\documentclass[final,english]{bullsrsl}[2022/06/15]

\usepackage[latin1]{inputenc}
\usepackage[T1]{fontenc}

\usepackage{natbib} 
\usepackage{graphicx}

\begin{document}
\title{Intranight variability of ultraviolet emission from high-$z$ blazars}%\\ \emph{Bulletin de la Société Royale des Sciences de Liège}}

%\author[affil={1}]{Krishan}{Chand}
\author[affil={}]{Krishan}{Chand}
%\author[affil={1}]{Lisa}{{\L}isander}
%\author[affil={2}]{Hum}{Chand}
%\author[affil={4}]{Lotta}{Lothardis}
\affiliation[]{Aryabhatta Research Institute of Observational Sciences (ARIES), Manora Peak, Nainital $-$ 263002, India}
%\affiliation[2]{Department of Physics and Astronomical Science, Central University of Himachal Pradesh (CUHP), Dharamshala 176215, India}
%\affiliation[3]{Institute of Improbability, Impossible City}
%\affiliation[4]{Double Blind Testing Inc., Great Opportunities}
\correspondance{krishanchand007.kc@gmail.com}
\date{13th June 2023}

\maketitle

% \author[affil1]{FirstName (+ MiddleInitials if necessary)}{FamilyName}
% \author[affil2]{...}{}
% \equalcontribauthor[]{}{} % Maximum two --> counter
% \consortium[affil]{Consortium Name}
% With consortium: affiliation will be set to "See Appendix 1 for a full
% list of consortium members and their respective affiliations
% \affiliation[affil1]{...}
% \affiliationq[affil2]{...}

% \correspondence[]{}
% No explicit corresponding author: use first author
% 

% Abstract of the paper in the same language as the paper
\begin{abstract}
Rapid intranight variability of continuum and polarization in blazars is a very useful tool to probe the beaming of a relativistic jet and the associated population of the relativistic particles. Such intranight variability in the rest-frame optical continuum has been carried out extensively, but there is a scarcity of such information in the rest-frame ultra-violet (UV), where the cause of variability might be due to {a} secondary population of relativistic particles. To fill this gap, recently in \citet{Krishan2022MNRAS.511L..13C} we reported for the first time intranight variability study of a sample of high-$z$ blazars so that the monitored optical radiation is their rest-frame UV radiation. Here we discuss in detail the implication of this investigation with a proper comparison of high-$z$ blazar samples with fractional optical polarization ($p_{opt}$) smaller and higher than 3\%. In this context, we also report intranight variability study of an additional high-$z$ blazar at $z$=2.347, namely J161942.38+525613.41, monitored over three sessions each with a duration of $\sim$5 hr. Our investigation does not reveal any compelling evidence for a stronger intranight variability of UV emission for high polarization blazars, in contrast to the blazars monitored in the rest-frame blue-optical. We also discuss this trend in light of the proposal that the synchrotron radiation of blazar jets in the UV/X-ray regime may arise from a relativistic particle population different from that radiating up to near-infrared/optical frequencies. 
\end{abstract}

\keywords{galaxies: active -- galaxies: photometry -- galaxies: high-redshift -- galaxies}

%\section{Section -- Level 1 title (Times New Roman, bold, 14 pts)}
\section{Introduction}
\label{intro}
Blazars are radio-loud active galactic nuclei (AGN) whose radiation is mostly non-thermal. The origin of radiation is a relativistic jet that is roughly directed and hence beamed towards the observer \citep{Blandford1978PhyS...17..265B,Moore1984ApJ...279..465M,Urry1995PASP..107..803U}. The radiation appears highly Doppler boosted in our direction due to beaming and hence completely dominates the radiation coming from the host galaxy and accretion disk. Rapid variability of continuum and polarized emission, and a flat radio spectrum (i.e., a dominant radio core) are well-known characteristics of beaming. The quasar subset of blazars with a higher power is known as flat-spectrum radio quasars {(FSRQs)} \citep[e.g.,][]{Stockman1984ApJ...279..485S,Wills1992ApJ...398..454W,Angelakis2016MNRAS.463.3365A}.  Their jets frequently show superluminal motion as well as large variability amplitude at all wavelengths from radio to $\gamma$-rays. The flux variability from radio to optical, and even TeV bands has been very useful for unveiling the physics of blazars, particularly the extreme cases of flux variability observed on hour-like or even shorter time-scales \citep[e.g.,][]{Wagner1995ARA&A..33..163W,Aharonian2017ApJ...841...61A,Gopal2018BSRSL..87..281G}. \par
The UV band is one important spectral region for which little information about intranight variability is available. The primary reason for the lack of information about the rapid UV variability of AGN is that intranight monitoring campaigns for space-borne UV telescopes are prohibitively time expensive. Although, in recent years, {there have been attempts to fill this information gap} by using UV data acquired with GALEX for large AGN samples{;} however, such studies have covered only day-like or longer time-scales and are particularly devoted to optically selected quasars \citep[e.g.,][]{Punsly2016ApJ...830..104P,Xin2020MNRAS.495.1403X}. {Hence, these studies may well have had substantial contributions from the accretion disk and not have measured only UV emission of synchrotron origin.}  Additionally, from detailed measurements of spectral energy distribution (SED) of a few quasar jets, a hint regarding UV radiation has emerged, according to which a very substantial, if not dominant contribution to synchrotron UV radiation of quasar jets may arise from a relativistic particle population distinct from the one responsible for their radiation up to near-infrared and optical frequencies. Specifically, the SED of the knots in the kiloparsec-scale radio jets of some quasars is found to exhibit a sharp spectral {\it upturn} towards the UV and connecting smoothly  thereafter to the X-ray data points \citep{Uchiyama2006ApJ...648..910U,Uchiyama2007ApJ...661..719U,Jester2007MNRAS.380..828J}. Moreover, the observed high polarization in the UV emission supports a synchrotron interpretation for this higher-energy radiation component \citep{Cara2013ApJ...773..186C}. The above-mentioned studies support the idea that the optical and UV radiations from jets are synchrotron radiation arising from two different populations of relativistic particles. This idea was recently reinforced by \citet{Krishan2022MNRAS.511L..13C} by comparing the intranight flux variability of blazars at rest-frame UV and optical wavelengths. \par
%The importance of looking for independent manifestations of this putative relativistic particle population has been underscored in \citet{Gopal-krishna2019BSRSL..88..132G}.
 In order to probe the rapid UV variability of blazars, the practical approach adopted by \citet{Krishan2022MNRAS.511L..13C} is that they carried out {\it optical} intranight monitoring of 14 blazars (FSRQs), located at very high redshifts (1.5 $<$ $z$ $<$ 3.7), so that their monitored optical radiation is actually UV emission in the rest-frame. Their study consisted of two prominent subclasses of blazars, distinguished by low and high fractional polarization measured in the optical, with division taken at $p_{opt} = 3\%$ \citep{Moore1984ApJ...279..465M}. This resulted in (i) nine low-polarization FSRQs with $p_{opt}<$3\% and five high-polarization FSRQs with $p_{opt}>$3\%. Based on the photometric statistical analysis for nine low-polarization FSRQs and five high-polarization FSRQs, they found no evidence for a strong correlation of intranight variability of UV emission with polarization, in contrast to the strong correlation found for intranight variability of optical emission. This led them to a proposal that the synchrotron radiation of blazar jets in the UV/X-ray regime arises from a relativistic particle population distinct from the one responsible for their synchrotron radiation up to near-infrared/optical frequencies. Since their two samples are too small to be representative of the high-$z$ blazar population, this finding might be spurious. Hence, an independent check on this finding is required which can be achieved through intranight optical monitoring of a larger  sample of high-$z$ blazars with low and high polarization. To make the above finding statistically  more robust, we made an enlarged sample of 34 high-$z$ blazars which is 2.4 times larger than the combined sample of low-polarization and high-polarization blazars reported in \citet{Krishan2022MNRAS.511L..13C} (see section \ref{sample}).  
 The major obstacle in enlarging the sample of high-$z$ blazars for studying intranight variability of rest-frame UV emission with polarization is the lack of polarization information in the literature. And it is essential to have high-$z$ blazars with low and high polarization to explore the dependence of intranight UV variability with polarization.   More than half of the  sources (19) in an enlarged sample of 34 high-$z$ blazars also lack polarization information in the literature; therefore, the purpose of the intranight variability study of UV emission for an enlarged sample of high-$z$ blazars is two-folded: (i) polarization measurements of the sources  and (ii) intranight optical monitoring of the sources.  The selected 34 high-$z$ blazars  have a brightness ($m_R$) range from 15.2 to 17.4 and to detect the polarization of a few per cent within this brightness range requires telescopes of at least 2-3 metres in diameter. Therefore, for polarization measurements, we are making proposals in consideration of the impending availability of an imaging polarimeter at the 3.6-m Devasthal Optical Telescope (DOT) at ARIES  and the intranight optical monitoring will be carried out using metre-class telescopes available in India.\par
  The polarization and intranight optical observations for high-$z$ blazar samples in \citet{Krishan2022MNRAS.511L..13C} are separated by a median time of four years. The fluctuation in the polarization state of around a quarter of FSRQs on a year-like time-scale has been reported in various studies \citep{Impey1990ApJ...354..124I,Krishan2022MNRAS.516L..18C,Krishan2023PASA...40....6C}. Owing to the variable nature of polarization for blazars, the two sets of observations i.e., polarimetric and photometric observations should be quasi-simultaneous.  However, it is quite challenging to conduct  both polarimetric and photometric observations simultaneously.  It should be noted that there exists a strong correlation between intranight optical variability (INOV) and $p_{opt}$ for moderately distant blazars despite the fact that the two sets of observations were made a decade apart \citep[][and references therein]{Goyal2012AA...544A..37G}.  It is also known that the long-term optical variability is stronger for sources with large $p_{opt}$ \citep[e.g.,][]{Angelakis2016MNRAS.463.3365A}.   
  % For an enlarged sample of 34 high-z blazars, we are proposing quasi-simultaneous polarimetric and photometric observations which we intend to obtain using the upcoming imaging polarimeter at 3.6-m DOT and photometric imagers at meter-class telescopes in India. This is a long-term project, so, here in this article, we present intranight optical monitoring for J161942.38+525613.41, one of the 34 high-$z$ blazars, in three sessions with a median duration of $\sim$ 5 hours.
   It will be useful not only to enlarge the sample of high-$z$ blazars for studying the intranight variability of the UV continuum but also to conduct intranight photometric observations in conjunction with their quasi-simultaneous polarimetric observations. This forms the main motivation of this article, where we devise a large sample of 34 high-$z$ blazars for intranight variability and quasi-simultaneous polarimetric observations based on the upcoming imaging polarimeter at 3.6-m DOT. In this ongoing long-term project, we present the intranight optical monitoring of one of the 34 high-$z$ blazars, namely J161942.38+525613.41 at $z=2.347$, in three sessions with a median duration of $\sim$5 {hr}.

The article is organised as follows. In section \ref{sample}, we present a selection of enlarged sample along with notes on the J161942.38+525613.41. Section \ref{monitoring}  details about photometric monitoring and data reduction, while statistical analysis is briefly given in section \ref{analysis}. Finally, we present our detailed  discussion and conclusion  in section \ref{discuss}.

\section{The enlarged sample of high-{\it z} blazars}
\label{sample}
The enlarged sample of high-$z$ blazars for intranight optical and polarimetric monitoring  was derived from the 3561 sources listed in the 5$^{th}$ edition of the ROMA blazar catalogue (ROMA-BZCAT, \citealp{Massaro2015ApSS.357...75M}). These sources are either confirmed blazars or exhibit blazar-like characteristics. The redshift requirement of $z> 1.5$ was first applied to 3651 sources, and 784 blazars were found to meet this criterion. Then, we applied a brightness filter i.e., $m_R <$ 17.5  to 784 blazars, to ensure a good SNR with metre-class telescopes. This resulted in 84 blazars, of which 41 were discarded due to their negative declinations. Nine additional sources were also omitted because a long intranight monitoring session was not possible as these sources transit at night during the monsoon period.  Thus,  the final enlarged sample of high-$z$ blazars consists of 34 sources with redshift $z >$ 1.5. Of the 34 blazars,  nine sources were already reported in \citet{Krishan2022MNRAS.511L..13C}. 
In this article, we have  provided the INOV results for one of the 34 high-$z$ blazars, namely, J161942.38+525613.41 along with the previously reported  results for blazars in \citet{Krishan2022MNRAS.511L..13C}  and the notes on this source are given below:\par
{\bf J161942.38+525613.41:} It is an FSRQ type of blazar with a brightness ($m_R$) of  16.7 mag \citep{Massaro2015ApSS.357...75M}. It is located at a redshift of 2.347 \citep{Herazo2021AJ....161..196P} with the logarithmic bolometric luminosity of 47.78 erg s$^{-1}$ \citep{SuvenduS2020ApJS..249...17R},  making it a member of the high-luminosity tail of blazars. The source has detection in radio bands having a flux density of 182 mJy and 128 mJy at 1.4 GHz \citep{Condon1998AJ....115.1693C} and 5 GHz \citep{Gregory1996ApJS..103..427G} respectively. A flat radio spectral index ($\alpha_r$) of $-$0.07 \citep{Bourda2010A&A...520A.113B}, signifies the core dominance of the source.  Due to its high-$z$ of 2.347, the monitored optical radiation is actually rest-frame UV emission and \citet{SuvenduS2020ApJS..249...17R} estimated  a  logarithmic continuum luminosity of 47.19 erg s$^{-1}$ and 46.84 erg s$^{-1}$ at 1350 {\AA} and 3000 {\AA} respectively.  

\section{Photometric Monitoring and Data Reduction}
\label{monitoring}
We carried out the intranight photometric observations of the high-$z$ FSRQ J161942.38+525613.41 ($z$=2.347) in Johnson-Cousins R or SDSS r-band in three sessions with a median duration of 5.06 {hr}. The 1.04-m Sampuranand Telescope (ST; \citealt{Sagar1999CSci...77..643S}, two sessions) and the 3.6-m Devasthal Optical Telescope (DOT; \citealt{Kumar18}, one session) are the telescopes that were employed. The observations for the ST sessions were taken with 4k $\times$ 4k CCD fixed at the focal plane of ST which is of Ritchey Chretien (RC) type \citep{Sagar1999CSci...77..643S}. The 4k $\times$ 4k CCD were cooled to $-120^{\circ}$C using liquid nitrogen to minimise the dark noise. The CCD has a pixel size of 15 microns and a plate scale of 0.23 arcsec per pixel, thus covering a field of view (FOV) of 15 $\times$ 15 arcmin$^2$ on the sky. The observations were conducted in 4 $\times$ 4 binning mode at a gain of 3e$^-$ per ADU and a readout speed of 1 MHz with a readout noise of 7e$^-$. For the DOT session, the observations were acquired with the ARIES Devasthal-Faint Object Spectrograph and Camera (ADFOSC) mounted at the Cassegrain main port of the DOT which is also of Ritchey Chretien (RC) design \citep{Kumar18}. ADFOSC is equipped with a deep depletion 4k $\times$ 4k CCD, cooled to $-120^{\circ}$C using a closed-cycle cryo cooling thermal engine having a pixel size of 15 microns, a plate scale of $\sim$ 0.2 arcsec per pixel, hence covering a 13.6 $\times$ 13.6 arcmin$^2$ FOV on the sky \citep{Omararticle}. The observations were also carried out in 4 $\times$ 4 binning mode with a readout noise of 8e$^-$ at a speed of $\sim$0.2 MHz and a gain of 1e$^-$ per ADU.\par
The pre-processing of the raw images (bias subtraction, flat-fielding and cosmic ray removal) was done using the standard tasks
available in the Image Reduction and Analysis Facility (IRAF). The instrumental magnitudes of the target source and the selected three comparison stars in the CCD frames were determined through aperture photometry \citep[]{Stetson1987PASP...99..191S,1992ASPC...25..297S}, using the Dominion Astronomical Observatory Photometry II (DAOPHOT II)
algorithm. The aperture radius, a very crucial parameter for photometry was determined by averaging the FWHM of the point spread function (PSF) of five moderately bright stars. In the present case, we fixed the aperture radius equal to 2 $\times$ PSF since the SNR was found to be the highest for this aperture (see \citealp{Krishan2022MNRAS.511L..13C}). We then derived differential light curves (DLCs) for the target source relative to the three selected comparison stars in the same CCD frames which are within 1.5 magnitudes of the target source (see Table~\ref{comparison_color1}). 

\begin{figure}
\centering
\includegraphics[scale=0.8]{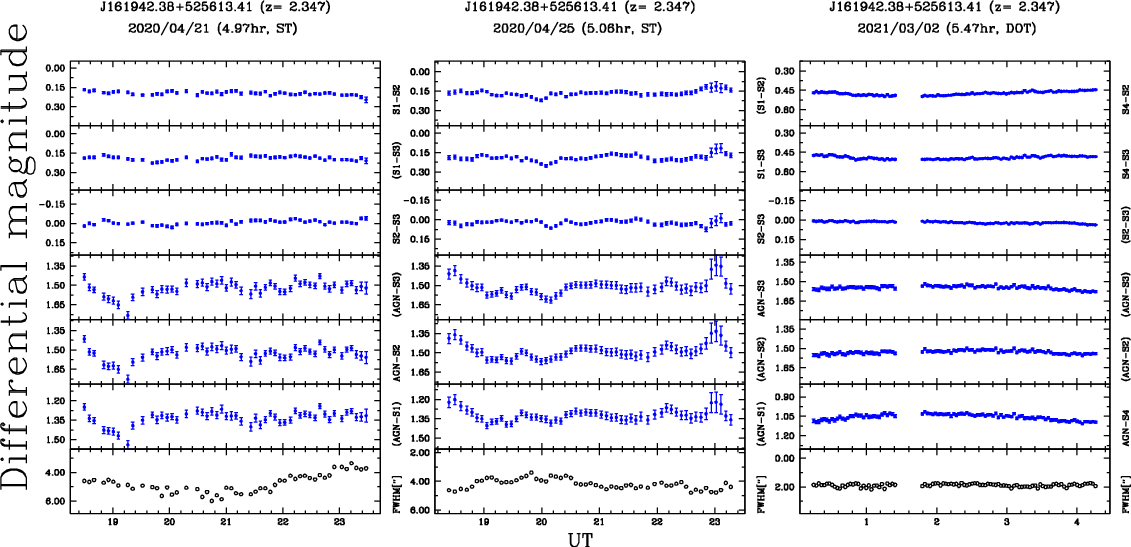}
\bigskip

\begin{minipage}{12cm}
\caption{Differential light curves (DLCs) for the high-$z$ FSRQ, J161942.38+525613.41. The date and duration of the observation along with the name of the source J161942.38+525613.41 are given at the top of each night's data. The upper three panels give the comparison star-star DLCs, whereas the subsequent lower panels give the source-star DLCs as described by the labels on the right side. The bottom panel gives the variation in seeing during the course of observation.}
\label{fig}
\end{minipage}
\end{figure}

\begin{table}
\scriptsize
\centering
\begin{minipage}{150mm}
\caption{Basic parameters of selected comparison stars for high-$z$ J161942.38$+$525613.41.}
\label{comparison_color1}
\end{minipage}
\bigskip

\begin{tabular}{ccc ccc c}
\hline

{Object  } &   Date       &   {R.A.(J2000)} & {Dec.(J2000)}                      & {\it g} & {\it r} & {\it g-r} \\
           &  yyyy/mm/dd    &   (hh:mm:ss)       &($^\circ$: $^\prime$: $^{\prime\prime}$)   & (mag)   & (mag)   & (mag)     \\
{(1)}      & {(2)}        & {(3)}           & {(4)}                             & {(5)}   & {(6)}   & {(7)}     \\
\hline

J161942.38$+$525613.41 & 2020/04/21, 2020/04/25, 2021/03/02 & 16:19:42.38&$+$52:56:13.41 &16.82  &16.74   &0.08   \\
S1 	   & 2020/04/21*, 2020/04/25*              & 16:19:32.72&$+$52:49:35.19 &16.41  &15.53   &0.88   \\
S2 	   & 2020/04/21, 2020/04/25*, 2021/03/02*   & 16:19:48.69&$+$52:58:19.60 &16.05  &15.37   &0.68   \\
S3     & 2020/04/21*, 2020/04/25, 2021/03/02*   & 16:19:37.29&$+$52:58:44.68 &15.92  &15.33   &0.59\\
S4     & 2021/03/02                         & 16:20:14.60&$+$52:52:17.80 &16.36  &15.88   &0.48\\
\hline
\multicolumn{7}{l}{The comparison stars on a particular marked session {($^{*}$)} have been used to determine the INOV status of AGN.}
%\hline
\end{tabular}
\end{table}

\section{Statistical Analysis}
\label{analysis}
In the current analysis, we have employed the $F_{\eta}$ test to determine whether INOV exists in the derived DLCs or not \citep{Krishan2022MNRAS.511L..13C,Gopal2023MNRAS.518L..13G}.  We further identified the two most stable comparison stars out of the three that were initially selected for a given session and applied the $F_{\eta}$ test to the three DLCs involving only them (the selected two comparison stars for each session are shown within  parentheses in column 5 of Table \ref{tab2} and the corresponding DLCs are also labelled within parentheses in Fig. \ref{fig} on the right side). In several independent studies, it has been found that the photometric errors returned by DAOPHOT are too small \citep{Gopal-Krishna1995MNRAS.274..701G,Garcia1999MNRAS.309..803G,Sagar2004MNRAS.348..176S,Stalin2004MNRAS.350..175S,Bachev2005MNRAS.358..774B,Goyal2012AA...544A..37G,GoyalA2013MNRAS.435.1300G} and the best value for the underestimation factor $\eta$ is found to be $1.54\pm0.05$, based on  262 monitoring sessions of quasars/blazars \citep{Goyal2013JApA...34..273G} and the same value has been used in this study. The F-values for the selected two blazar DLCs in a given session are:
%of F$-\eta$ is written as:
\begin{equation} 
\label{eq.ftest2}
F_{1}^{\eta} = \frac{Var(q-s1)}
{ \eta^2 \sum_\mathbf{i=1}^{N}\sigma^2_{i,err}(q-s1)/N},  \\
\hspace{0.1cm} F_{2}^{\eta} = \frac{Var(q-s2)}
{ \eta^2 \sum_\mathbf{i=1}^{N}\sigma^2_{i,err}(q-s2)/N}  \\
\end{equation}\\
where $Var(q-s1)$ and $Var(q-s2)$ are the variances of the target source relative to  star1 and star2 respectively. And $\sigma^2_{i,err}(q-s1)$ and $\sigma^2_{i,err}(q-s2)$ denote the errors that DAOPHOT returned for each individual data point of the target source relative to star1 and star2 respectively. N is the total number of data points taken in observation (Column 3 of Table~\ref{tab2}) and $\eta= 1.54$ is the scaling factor, as mentioned above. The computed value of $F_{1}^{\eta}$ and $F_{2}^{\eta}$ for the target source$-$star DLCs are given in Table~\ref{tab2}, column 5.\par
The F$-$values estimated from the F$_{1,2}^{\eta}$ test (Column 5 from Table~\ref{tab2}) are compared with the critical value
of F ($F_{c}^{\alpha}$) for $\alpha = $ 0.05, 0.01 corresponding to confidence levels of 95\% and 99\%, respectively (Column 6,7 in Table~\ref{tab2}). If the computed $F-$value exceeds the critical value, the null hypothesis (i.e., no variability) is discarded. We thus classify a source with F$-$value $\ge$ F$_{c}$(0.99) as a variable (`V') at confidence level $\ge$0.99; probably variable (`PV') for $F-$value between F$_{c}$(0.95) and F$_{c}$(0.99), and non-variable  (`NV') if the computed $F-$value is less than F$_{c}$(0.95). In the F$_{\eta}$ test where two F$-$values involved are related to comparison stars 1 and 2, we set the class of a target source as variable (`V') if F$-$value calculated for both source-star1 and source-star2 DLCs exceeds the value F$_{c}$(0.99). And the target source is considered as a non-variable (`NV') if  F$-$value for at least one of the source-star DLCs is less than the F$_{c}$(0.95). For other remaining situations i.e., a combination of V, PV or PV, PV, the target source is classified as a probable variable (`PV').  Column 10 in Table~\ref{tab2} lists the `Photometric Noise Parameter' (PNP) = {$\sqrt { \eta^2\langle \sigma^2_{i,err} \rangle}$ }, estimated for a monitoring session using the star-star DLCs, where ${\eta=1.54}$, as mentioned above. \par
{To make the variability results more convincing, we attempted to derive the DLCs using the same set of three comparison stars (S1, S2 \& S3) for the two ST and one DOT sessions observed on April 21, 2020, April 25, 2020, and March 02, 2021, respectively. However, due to the lack of steadiness of S1 for the DOT session observed on March 02, 2021, a new stable comparison star S4 was used instead of S1. Since the F$_{\eta}$ test incorporates two comparison stars, we further identified the two most stable comparison stars out of the three for the observed sessions as mentioned above. We found a set of stable comparison stars (S1, S2 \& S3) in different combinations for the two ST and one DOT sessions (see Table \ref{tab2}, also, Fig. \ref{fig}). If we look at Fig. \ref{fig}, it is easy to believe that there are significant variations on the first night and a good chance that there are some on the second night, but there is little chance of real variability on the third night, except perhaps for a slow rise and fall. However, based on the F$_{\eta}$ test, Table \ref{tab2} indicates that the first is PV, the second is NV, and the last is V. Presumably, this non-obvious result comes from some combination of the larger number of data points that could be measured in the same amount of time at the larger DOT than at ST as well as the smaller error bars at DOT.}\par
The INOV variability  amplitude ($\psi$, see Table~\ref{tab2}, column 11) was computed using the definition given by \citep{Heidt1996A&A...305...42H}:\\
 $\psi= \sqrt{({A_{max}}-{A_{min}})^2-2\sigma^2}$\\
Here $A_{max}$ and $A_{min}$ are the maximum and minimum values in the source-star DLC and $\sigma^2=\eta^2<\sigma^2_{q-s}>$, where, $\sigma^2_{q-s}$ is the mean square rms error for the data points in the DLC and ${\eta=1.54}$. The mean value of $\psi$ for a session, i.e., the average of the $\psi$ values estimated for the two selected DLCs of the target source is given in column 11 of Table \ref{tab2}.%\par 

%\subsection{Estimation of Duty Cycle} 
%For computing the duty cycle (DC) of INOV, we used the definition given by \citet{Romero1999A&AS..135..477R}:
%\begin{equation} 
%DC = 100\frac{\sum_\mathbf{i=1}^\mathbf{n} N_i(1/\Delta t_i)}{\sum_\mathbf{i=1}^\mathbf{n}(1/\Delta t_i)} {\rm \%} 
%\label{eq:dc} 
%\end{equation}
 %Here $\Delta t_i = \Delta t_{i,obs}(1+z)^{-1}$ is the intrinsic duration, corrected for its cosmological redshift, $z$, and $\Delta t_{i,obs}$ is the duration over which a source on an $i^{th}$ night has been observed (Column 4 in Table~\ref{tab2}). Since the duration of observation for  a particular source is not uniform on different nights, the formula of DC is weighted by the observed monitoring duration $\Delta t_i$ on an $i^{th}$ night. Here, $N_i$ was taken as 1 if INOV was detected, otherwise $N_i$ = 0.\\

\begin{table}
\scriptsize
\centering
\begin{minipage}{170mm}
\caption{Result of the statistical test for detecting INOV in the DLCs of the  J161942.38+525613.41.}
\label{tab2}
\end{minipage}
\bigskip

\begin{tabular}{ccccccccccccc}
%\begin{tabular}{cccccccccccl}  
  \hline 
  %\centering
\multicolumn{1}{c}{Blazar}
&\multicolumn{1}{c}{Date}
&{N}
&{T} 
&\multicolumn{3}{c}{F-test}
 &\multicolumn{2}{c}{INOV}
 
 &\multicolumn{1}{c}{$\sqrt { \eta^2\langle \sigma^2_{i,err} \rangle}$}
 & $\overline\psi$ &\\%{$ \sigma_{s1-s2}$}\\
  (SDSS name)       & yyyy/mm/dd & &{(hr)} &{$F_{1}^{\eta}$},{$F_{2}^{\eta}$}&{$F_{c}(0.95)$}&{$F_{c}(0.99)$}&\multicolumn{2}{c}{status$^{a}$} && \%&\\
 (1)     &   (2)    &(3)&(4)  & (5)                         &(6)       &(7)           &(8)            &(9)     &(10)     & (11)     \\%%
\hline
                                                                                                
 J161942.38+525613.41 &21/04/2020  &51&4.97    &  1.83 (S1),   1.78 (S3)&  1.60&  1.95& PV      ,     PV  &PV     & 0.015 (S1-S3) &29.60\\
 J161942.38+525613.41 &25/04/2020 &58&5.06       &  0.60 (S1),   0.62 (S2)&  1.55&  1.87&NV      ,     NV  &NV     & 0.037 (S1-S2)&--- \\
 J161942.38+525613.41 &02/03/2021 &138&5.47      &  6.63 (S2),   9.62 (S3)&  1.33&  1.49& V      ,      V   & V    & 0.005 (S2-S3)&15.89 \\
   \hline
\multicolumn{11}{l}{$^a$ V=variable, i.e., confidence
       $\ge 0.99$; PV = probable variable ($0.95-0.99)$; NV
       = non-variable ($< 0.95$).}\\
\multicolumn{11}{l}{Variability status identifiers (col. 8), based on AGN-star1 and AGN-star2 DLCs are separated by a comma.}
\end{tabular}
\end{table}

\section{Discussion and Conclusion}
\label{discuss}
The R-band/r-band intranight monitoring reported in \citet{Krishan2022MNRAS.511L..13C} for the two samples of high-$z$ blazars corresponds to their rest-frame UV emission. For sources in their first sample (nine $LP_{FSRQs}$), the rest-frame wavelengths of monitoring are from 1374 {\AA} to 2078 {\AA} (median 1973 {\AA}), while for their second sample (five $HP_{FSRQs}$), the corresponding range is from 2078 {\AA} to 2495 {\AA} (median 2099 {\AA}). The rest-frame wavelength of the high-$z$ FSRQ J161942.38+525613.41, located at a redshift of 2.347 is 1914 {\AA}, thus making it a suitable source for studying the intranight variability of UV emission.  

\citet{Krishan2022MNRAS.511L..13C} estimated the duty cycle (DC) of $\sim$ 30\% with $\psi > 3\%$ for the intranight variability of nine $LP_{FSRQs}$ at median $z \sim$ 2.25. For five $HP_{FSRQs}$ at median $z \sim$ 2.0, the DC of intranight variability was estimated to be $\sim12\%$ with $\psi > 3\%$. On two counts, the results of $F_{\eta}$ test for nine $LP_{FSRQs}$ and five $HP_{FSRQs}$ are different from the past INOV estimates using the same $F_{\eta}$ test. Firstly, \citet{GoyalA2013MNRAS.435.1300G} estimated an INOV DC of $\sim$ 10\% for $\psi > 3\%$ cases, using a large sample of 12 moderately distant (median $z \sim 0.7$) $LP_{FSRQs}$ monitored in 43 intranight sessions of $>$ 4 hr duration. In the same study, the INOV DC was found to be $\sim$ 38\% for $\psi > 3\%$, for a sample of 11 moderately distant (median $z \sim 0.7$) $HP_{FSRQs}$ monitored in 31 sessions.  It should be noted that based on these results, a strong correlation between intranight optical variability (INOV) and $p_{opt}$  was established for moderately distant blazars despite the fact that the two sets of observations were made a decade apart. This indicates that the propensity of a given FSRQ to exhibit strong INOV is of a fairly stable nature and it correlates tightly with the optical polarization class.\par  %It is also known that the long-term optical variability is stronger for sources with large $p_{opt}$ \citep[e.g.,][]{Angelakis2016MNRAS.463.3365A}.\par
The DC of $\sim$ 30\% for high-$z$ $LP_{FSRQs}$ is considerably higher than that of their lower-$z$ counterparts (DC $\sim$ 10\%) and is quite comparable to their lower-$z$ $HP_{FSRQs}$. The excess of DC found for high-$z$ $LP_{FSRQs}$ (DC $\sim$ 30\%) could possibly be even greater given that their sample of $LP_{FSRQs}$ is heavily biased towards the most luminous members of this AGN class and variability is known to anti-correlate with luminosity from several studies \citep[e.g.,][]{Guo2014ApJ...792...33G}. However, it is still unknown whether the anti-correlation with luminosity is significant on the intranight time scales since so far, it has been established only for variability on month/year-like time scales. One possible explanation offered by \citet{Krishan2022MNRAS.511L..13C} for the high DC of $LP_{FSRQs}$ is that some of their members are actually $HP_{FSRQs}$ but were mistakenly categorised as $LP_{FSRQs}$ since their observed polarization, $p_{opt}$ (i.e., rest-frame UV polarization) has been decreased due to dilution by thermal UV from the accretion disc.\par
Secondly, their second sample of five $HP_{FSRQs}$ did not support the above explanation. They estimated an INOV DC of $\sim$12\% for five $HP_{FSRQs}$ which is quite low{,} and these $HP_{FSRQs}$ can not be misidentified as $LP_{FSRQs}$ since thermal dilution only lowers the polarization. Furthermore, \citet{Krishan2022MNRAS.511L..13C} looked for the observational biases (i.e., the intrinsic duration of monitoring and photometric sensitivity) that could have spuriously led to higher DC for $LP_{FSRQs}$ in comparison to $HP_{FSRQs}$. They found that the intrinsic duration of monitoring and photometric sensitivity for $LP_{FSRQs}$ and $HP_{FSRQs}$  is very similar. Hence, the possibility of observational biases was discarded. In summary, based on the results for nine $LP_{FSRQs}$ and five $HP_{FSRQs}$, a strong correlation of intranight variability of UV emission with polarization could not be found, in contrast to the strong correlation found for intranight variability of optical emission. \par 
To this astonishing result, we now added three intranight sessions of FSRQ J161942.38+525613.41 located at $z=2.347$, monitored in R-band/r-band with a median duration of $\sim$5 hr using ST and DOT.  Since the polarization information is unavailable in the literature, the FSRQ J161942.38+525613.41 could not be classified as $LP_{FSRQ}$ or $HP_{FSRQ}$. Therefore, the FSRQ  J161942.38+525613.41 was included in both the samples of \citet{Krishan2022MNRAS.511L..13C} while estimating the  INOV DC. For estimating the INOV DC, only the type `V' sessions with confirmed variability of $\psi$ $>$ 3\% were used. Only one session was found to be variable (`V') with $\psi$ $>$ 3\%, observed on March 02, 2021 using DOT (see, Column 9 of Table \ref{tab2}). A high DC of $\sim$ 30\% with an amplitude of $\psi>$ 3\% was estimated for intranight variability of low-polarization FSRQs even after including the FSRQ J161942.38+525613.41 as a low-polarization FSRQ. Similarly, when the source J161942.38+525613.41 was taken into account as a high-polarization FSRQ, the estimated DC was found to be $\sim$ 16\% with an amplitude of $\psi>$ 3\% which is still fairly low. Thus, we found no evidence for a strong intranight variability of UV emission with polarization even after including the FSRQ J161942.38+525613.41 as a low/high-polarization FSRQ, in contrast to the blazars monitored in the rest-frame blue-optical. This supports the proposal put forth in \citet{Krishan2022MNRAS.511L..13C} that the synchrotron radiation of blazar jets in the UV/X-ray regime arises from a relativistic particle population distinct from the one responsible for their synchrotron radiation up to near-infrared/optical frequencies.  However, the measurement of  polarization  for the FSRQ J161942.38+525613.41 using the upcoming imaging polarimeter at 3.6-m DOT will make the source either $LP_{FSRQs}$ or  $HP_{FSRQs}$, resulting in an increment of  the size of one of two blazar samples of \citet{Krishan2022MNRAS.511L..13C} by just one which is again small. The two blazar samples are yet {too} small to be representative of the high-$z$ blazar population, {so} the above finding might be spurious. Hence, this interesting finding requires independent confirmation through intranight optical monitoring of a larger sample of high-$z$ blazars with low and high polarization. Consequently, an enlarged sample of 34 high-$z$ blazars was derived from the ROMA-BZCAT catalogue to verify the above finding which we intend to observe in future using metre-class telescopes available in India (see section \ref{intro}, also, section \ref{sample}).

\begin{acknowledgments}
{ We thank Prof. Paul J. Wiita for his valuable comments and suggestions for improving the quality of the manuscript.} We thank the organisers for allowing us to present our work at the BINA conference and for the local support. We also thank Prof. Hum Chand for his thoughtful suggestions and discussion regarding the current work. 
\end{acknowledgments}

\begin{furtherinformation}

\begin{orcids}
\orcid{0000-0002-6789-1624}{Krishan}{Chand}
%\orcid{1111-2222-3333-4444}{Leonie}{van Leon}
%\orcid{2222-3333-4444-5555}{Lotta}{Lothardis}

%{\sl This section is optional.
%You may list here the ORCIDs of those authors who would like to share them, one per line, with the \verb|\orcid{|\texttt{\emph{ORCID}}\verb|}{|\texttt{\emph{First name}}\verb|}{|\texttt{\emph{Last name}}\verb|}| command.
%This command typesets the information, and makes the ORCIDs themselves active links to the corresponding records on \href{https://orcid.org}{orcid.org}.

%Unlike in this sample, no other text should actually be included here and this section should reduce to a bare list.
%The \verb|\orcid| command controls line feeds by itself; please do not insert any \verb|\\| or \verb|\newline| before or after them.}
\end{orcids}

%\begin{authorcontributions}
%This section is mandatory when there is more than one author.
%The contributions of each author (identified by their initials) must be declared.
%We recommend to follow the \href{http://credit.niso.org}{CRediT} taxonomy (Contributor Roles Taxonomy).
%\end{authorcontributions}

\begin{conflictsofinterest}
The author declares no conflict of interest.
\end{conflictsofinterest}

\end{furtherinformation}

\bibliographystyle{bullsrsl-en}

\bibliography{references}

\end{document}